\documentclass[tighten,twocolumn]{aastex63}

\usepackage{graphicx}
\usepackage{ulem}
\usepackage{amsmath}
\usepackage{wrapfig}
\usepackage[thinlines]{easytable}
\usepackage{tikz}
\usepackage{hyperref}

\graphicspath{{./}{figs/}}

\begin{document}

\title{Shadows of the Colossus: Hierarchical Black Hole Mergers in a 10-million-body Globular Cluster Simulation}

\correspondingauthor{Aidan Mai}
\email{aimai@ucsd.edu}

\author[0009-0003-8963-7787]{Aidan Mai}
\affiliation{Department of Astronomy \& Astrophysics, University of California, San Diego; La Jolla, CA 92093, USA}

\author[0000-0002-4086-3180]{Kyle Kremer}
\affiliation{Department of Astronomy \& Astrophysics, University of California, San Diego; La Jolla, CA 92093, USA}

\author[0000-0003-4412-2176]{Fulya K{\i}ro\u{g}lu}
\affiliation{Center for Interdisciplinary Exploration \& Research in Astrophysics (CIERA) and Department of Physics \& Astronomy \\ Northwestern University, Evanston, IL 60208, USA}

\begin{abstract}
The LIGO/Virgo/Kagra (LVK) Collaboration has detected numerous binary black hole mergers with properties that challenge standard binary evolution scenarios, such as component masses above the pair-instability gap and high spin magnitudes. Dense stellar environments such as globular clusters provide a natural channel for producing such systems through hierarchical mergers, where black hole remnants formed in earlier mergers are retained in the cluster and undergo successive mergers. However, gravitational-wave recoil kicks often eject merger remnants from typical globular clusters, which limits hierarchical growth. Massive clusters with deeper potential wells, such as those found in giant elliptical galaxies like M87, may overcome this barrier, but direct simulations of such massive globular clusters remains computationally challenging. In this study, we present a 10-million-body cluster simulation performed with the \texttt{Cluster Monte Carlo} (\texttt{CMC}) code, referred to as \texttt{colossus}, which serves as a proxy for the most massive low-metallicity globular clusters observed in the local Universe. This simulation demonstrates that extended chains of hierarchical mergers can occur in massive globular clusters, producing black holes up to fifth generation with masses approaching $250\,M_\odot$, comparable to the most massive LVK events observed to date (e.g., GW231123). Combining the \texttt{colossus} simulation with the previous \texttt{CMC Cluster Catalog}, we develop a framework to extrapolate binary black hole merger predictions for the thousands of globular clusters seen in the Virgo Supercluster.
\vspace{1cm}
\end{abstract}

\section{Introduction}

Over the past decade, the LIGO/Virgo/KAGRA (LVK) network of gravitational wave (GW) detectors have been immensely successful in detecting binary black hole (BBH) mergers. Many of these GW detections have parameters that challenge typical binary star evolution scenarios, including black hole (BH) masses above the pair-instability mass gap, high spin magnitudes, and possible evidence for spin-orbit misalignment and high binary eccentricity. The recent event GW231123 \citep{GW231123}, with component masses of roughly $140\,M_{\odot}$ and $100\,M_{\odot}$ and high component spins of roughly $0.9$ and $0.8$, is a case in point. Events like GW231123 and other massive events may suggest that some of today's GW detections may have formed via channels alternative to classic stellar evolution pathways.

Recent work demonstrates the efficiency of dense stellar environments, e.g., globular clusters (GCs), young/open star clusters, and nuclear star clusters, in the dynamical formation of BBH systems \citep[e.g.,][]{PortegiesZwartMcMillan2000,Rodriguez2016mergers,AntoniniRasio2016,Askar2017,Banerjee2017,Samsing2018,Fragione2019,DiCarlo2019,kremer2020modeling,Ye2025}. These dense environments specifically enable repeated BH mergers, or hierarchical mergers, where BHs that formed in earlier BBH mergers are retained in the cluster and subsequently undergo additional mergers \citep[e.g.,][]{MillerHamilton2002,GerosaBerti2017,Rodriguez2019,Antonini2019,FragioneRasio2023}. This process provides a natural pathway for forming high-mass, high-spin BHs, and is commonly touted as a possible mechanism for forming the most massive LIGO events observed to date \citep[e.g.,][]{Fishbach2017,Kimball2021,GW231123}.

The main challenge preventing the growth of BHs through hierarchical mergers is the recoil kicks attained by BH merger products, due to anisotripic emission of GWs \citep[e.g.,][]{Bekenstein1973,Favata2004}. For typical mass ratios and spin values, these kicks often exceed $100\,$km/s, and in some cases can even reach $2000\,$km/s or more \citep[e.g.,][]{Merritt2004,Campanelli2007,Lousto2008,GerosaKesden2016}. For typical GCs with masses $10^5-10^6\,M_\odot$ and escape velocities less than roughly $50\,$km/s \citep[e.g.,][]{BaumgardtHilker2018}, the prompt ejection of BH merger products due to these velocity kicks generally inhibit hierarchical growth beyond one, or at most two, successive merger generations. Indeed, many previous studies performing $N$-body simulations of GCs have shown that BH mergers with at least one second generation (2G) component born from a previous merger constitute at most 10\% of all mergers in these systems, \citep[e.g.,][]{Rodriguez2019}.

However, in the most massive GCs and nuclear star clusters with higher escape velocities, BH merger products are more easily retained, enabling significantly more high-generation hierarchical mergers \citep[e.g.,][]{Rodriguez2020_SSC,FragioneRasio2023}. For example, the Milky Way's most massive GC, $\omega\,$Centauri, has an estimated mass of roughly $4 \times 10^6\,M_\odot$ and a (present-day) central escape velocity of roughly $60\,$km/s \citep{BaumgardtHilker2018}. \citet{Haberle2024} revealed an intermediate-mass BH of mass roughly $10^4\,M_{\odot}$ in the center of $\omega\,$Cen, which may have formed via repeated mergers with lower-mass BHs over the cluster's lifetime \citep{GonzalezPrieto2025}. Massive elliptical galaxies contain far more GCs relative to the Milky Way \citep[e.g.,][]{BrodieStrader2006}, including a much larger sample of very massive GCs. For example, M87 in the Virgo Supercluster is estimated to host in excess of $10^4$ GCs, including at least 100 GCs with masses in excess of $10^6\,M_{\odot}$, including a handful up to $10^7\,M_{\odot}$ \citep{Jordan2009}.  

Although the potential utility of these massive clusters in forming massive BHs via hierarchical mergers is clear, these systems pose a formidable challenge from a simulation perspective. The most massive direct N-body simulations contain up to $10^6$ stars, but have densities far lower than realistic systems and prohibitively long run times \citep[e.g.,][]{wang2016dragon}; however considerable recent progress has been made on this front with the \texttt{PETAR} code \citep{PETAR2020,BarberAntonini2025}. Even for Monte Carlo cluster methods which are considerably faster than direct N-body \citep[e.g.,][]{Askar2017,Rodriguez2022}, only a handful of massive simulations incorporating the most modern physics for BHs have been performed, often limited to high metallicities or shorter run times \citep[for example, see the ``\texttt{behemoth}'' model of][]{Rodriguez2020_SSC}. Due to these numerical challenges, many studies have adopted analytic or semi-analytic approaches to study the BBH mergers expected in these very massive systems \citep[e.g.,][]{AntoniniGieles2020,FragioneRasio2023,RAPSTER2024}. 

In this study, we present a 10-million body star cluster simulation performed with the \texttt{Cluster Monte Carlo} code, \texttt{CMC}. We refer to this model as ``\texttt{colossus}.'' Evolved for nearly $12 \,$Gyr and featuring modern treatments for all physics relevant to BHs and GW sources, \texttt{colossus} is, to our knowledge, the first of its kind published. With a final mass of roughly $4\times10^6\,M_{\odot}$ and metallicity of $0.1Z_{\odot}$, this simulation is intended as a proxy for the most massive low-metallicity GCs observed in massive elliptical galaxies like M87. Using this model, we demonstrate that massive BHs can form via extended chains of hierarchical mergers, with two of the most massive BHs formed in this model reaching fifth generation and masses approaching $250\,M_{\odot}$, comparable to the most massive LVK events observed to date, including GW231123.

This paper is organized as follows. In Section~\ref{sec:modeling}, we describe the \texttt{CMC} code and simulation details. In Section~\ref{sec:results}, we describe the main results from our \texttt{colossus} simulation, including the evolution of cluster properties and the BH population over time. In Section~\ref{sec:properties}, we compare the results of \texttt{colossus} with lower-mass models from the \texttt{CMC Cluster Catalog} \citep{kremer2020modeling}, confirming several trends in cluster mass related to BH mergers and ejected binaries derived in previous literature. In Section~\ref{sec:virgo}, we map the observed Virgo Supercluster GCs to their closest \texttt{CMC} counterparts using a mass-$r_h$ distance function, allowing extrapolation of \texttt{CMC} BBH results to the wider Virgo system as a whole. We conclude and discuss in Section~\ref{sec:summary}.

\section{Modeling Massive Star Clusters with CMC}
\label{sec:modeling}

To model our 10-million body cluster \texttt{colossus}, we use the \texttt{Cluster Monte Carlo} code \texttt{CMC} \citep{Joshi2000,Pattabiraman2013,Rodriguez2022}. \texttt{CMC} is a fully-parallelized H\'{e}non-type \citep{Henon1973} Monte Carlo code for stellar dynamics. This code includes all physics relevant to the evolution of dense star clusters, including two-body relaxation, tidal mass loss, stellar collisions and mergers, and direct integration of small-$N$ resonant encounters \citep{Fregeau2007} including post-Newtonian effects \citep{Rodriguez2018}. \texttt{CMC} is fully coupled to stellar evolution by implementing the publicly released \texttt{COSMIC} software \citep{Breivik2020}, which provides rapid calculations of single and binary star evolution with up-to-date physics for massive star evolution, interacting binaries, and compact object formation. Additionally, \texttt{CMC} output can be converted into various observational quantities, such as surface brightness profiles, velocity-dispersion profiles, and color-magnitude diagrams, using the \texttt{cmctoolkit} package \citep{Rui2021}. This enables robust comparisons with realistic GCs in both the Milky Way and other galaxies. The latest catalogs of \texttt{CMC} models \citep[e.g.,][]{kremer2020modeling} have proven remarkably successful in reproducing observed features of present-day GCs such as NGC~3201, NGC~6397, 47~Tuc, and others \citep{Kremer2019a,Rui2021, Vitral2022, Ye2022}. This has enabled detailed study of a range of compact object sources including radio pulsars \citep{Ye2019}, low X-ray binaries \citep{Kremer2018a}, white dwarfs \citep{Kremer2021_wd}, and binary BH mergers \citep{Rodriguez2016mergers}. For an in-depth description of the details of \texttt{CMC}, see \citet{Rodriguez2022}.

Our new $N=10^7$ model is intended to be an extension of the models in the \texttt{CMC Cluster Catalog}, therefore we adopt the same initial parameters and physical assumptions adopted there \citep[for further detail, see][]{kremer2020modeling}. We assume an initial virial radius $r_v=2\,$pc, initial metallicity $Z=0.1\,Z_{\odot}$, and a galactocentric radius $R_{\rm gc}=20\,$kpc in a Milky Way-like galactic potential. All initial stellar masses are drawn from a \citet{Kroupa2001} initial mass function ranging from $0.08-150\,M_{\odot}$, yielding a total initial mass of $6.0\times10^6\,M_{\odot}$. We assume an initial binary fraction of $5\%$ across all stellar mass, with binary mass ratios drawn uniformly in the range $q \in [0.1,1]$ and binary orbital periods drawn uniformly in log-space from near contact to the local hard-soft boundary. BH masses are computed using the fallback prescriptions of \citet{Fryer2012} and the (pulsational) pair-instability prescriptions of \citet{Belczynski2016}. Following \citet{FullerMa2019spin}, we assume all first generation BHs formed via stellar collapse are born with zero spin ($\chi=0$); we discuss this assumption further in Section~\ref{sec:hierarchical}. We compute spins, masses, and recoil kicks of binary BH merger products followings the method described in \citet{Rodriguez2018}, which implements  fits to numerical and analytic relativity calculations \citep{Campanelli2007,Gonzalez2007,BarausseRezzolla2009,LoustoZlochower2013,GerosaKesden2016}. We run the simulation for roughly $12\,$Gyr.


\section{Results}
\label{sec:results}

\subsection{Evolution of cluster parameters}

\begin{figure}
    \centering
    \includegraphics[width=\linewidth]{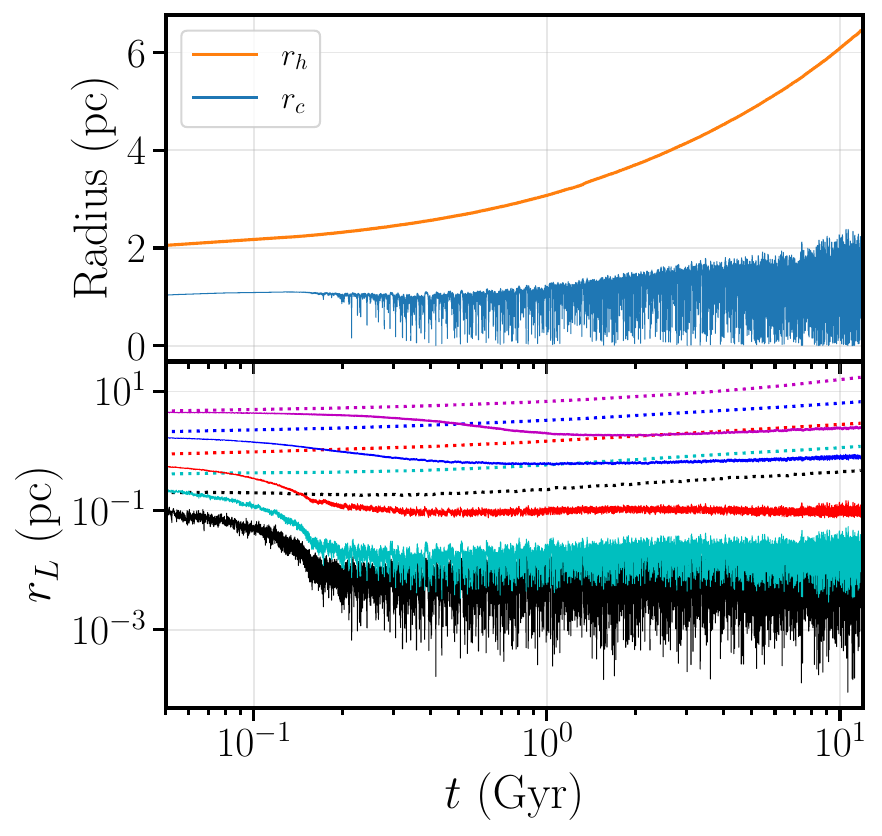}
    \caption{Top panel: Half-light radius ($r_h$) and core radius ($r_c$) of \texttt{colossus} plotted over time. Bottom panel: Lagrange radii of \texttt{colossus} plotted over time, calculated separately for black holes (solid curves) and stellar objects (dotted lines). Each Lagrange radii curve encloses a fixed fraction of mass for the particular type of object: from bottom to top, 0.1\%, 1\%, 10\%, 50\%, and 90\%. Stellar objects steadily expand in radius over time, while black holes segregate towards the core; by 2 Gyr, the 90\% BH threshold is smaller than the 10\% stellar threshold.}
    \label{fig:radius}
\end{figure}

The half-mass radius ($r_h$), core radius ($r_c$) and Lagrange radii ($r_L$) of \texttt{colossus} are plotted over time in Figure~\ref{fig:radius}. The core radius is defined using the density-weighted definition of \cite{CasertanoHut1985}, which features many sharp spikes indicative of core collapse episodes of the small-$N$ black hole subcluster \citep[e.g.,][]{BreenHeggie2013}. These black hole collapses are key to the formation of black hole binaries via three-body encounters \citep[e.g.,][]{Morscher2015}, which ultimately dynamically heat the cluster as a whole \citep{Kremer2020_bhburning}. We display Lagrange radii separately for the black hole (solid lines) and non-black hole populations (dashed lines). This distinction demonstrates the decoupling of the black holes from the stars on a time scale of roughly $100\,$Myr \citep[the mass segregation time of the black holes; e.g.,][]{Kremer2026_review}, as well as the subsequent core-collapse episodes of the black hole subsystem shown in the core radius curve in the top panel.

\begin{figure}
    \centering
    \includegraphics[width=\linewidth]{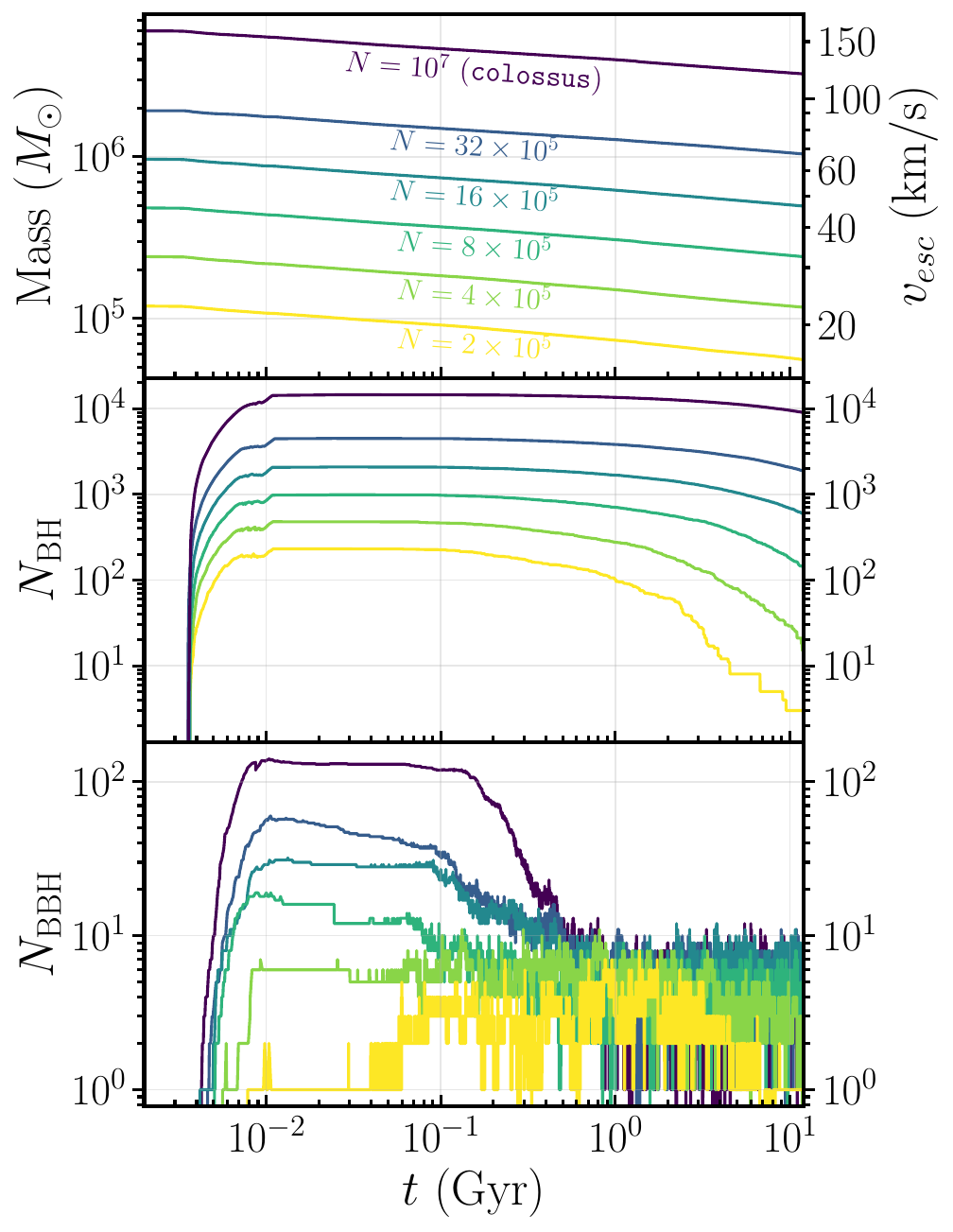}
    \caption{Evolution of cluster mass (top panel), number of BHs ($N_{\rm BH}$, center panel), and number of BBHs ($N_{\rm BBH}$, bottom panel) over time for \texttt{colossus} and five other models with varying initial mass. }
    \label{fig:nbh}
\end{figure}

The mass, number of black holes, and number of BH binaries over time in \texttt{colossus} are all plotted in Figure~\ref{fig:nbh}, alongside five other clusters from the \texttt{CMC Cluster Catalog} with the same initial parameters as \texttt{colossus}, except for initial cluster mass. In \citet{kremer2020modeling}, these are models 101, 102, 103, 104, and 146.\footnote{Note that model 146 in \citet{kremer2020modeling} has metallicity $Z=0.01Z_{\odot}$, different from $Z=0.1Z_{\odot}$ adopted for \texttt{colossus} and models 101-104. We did not run an $N=32\times10^5$ at $Z=0.1Z_{\odot}$ as part of the \texttt{CMC Cluster Catalog}. However, as shown in \citet{kremer2020modeling}, the distinction between these two low metallicities makes little difference in the BH mass distribution nor the overall cluster evolution.} The mass-loss histories of all six models (top panel of Figure~\ref{fig:nbh}) follow the same characteristic shape: a rapid early decline due to stellar evolutionary processes (supernovae, stellar winds, etc.), followed by a more gradual decline dominated by two-body relaxation, tidal stripping, and strong encounters in the cluster core \citep[e.g.,][]{Weatherford2023EscapeMech}. As a result, all six mass-loss curves are nearly identical in shape, differing only in a vertical offset corresponding to initial cluster mass.

A comparable trend can be observed in the models' BH population curves (middle panel of Figure~\ref{fig:nbh}). Since all six models utilized the same stellar initial mass function, the number of black holes throughout the end of the massive star evolution phase ($t \lesssim 10\,$Myr) scales directly with the initial cluster mass, producing identically-shaped BH population curves at early timescales. However, at later timescales where relaxation processes dominate, the BH loss rate becomes dependent on the cluster's relaxation timescale, which scales with cluster mass \citep[e.g.,][]{Morscher2015}. Therefore, lower-mass clusters having shorter relaxation times eject their black holes more rapidly, while higher-mass clusters retain a significant BH population for much longer. \texttt{colossus} retained nearly two-thirds of its initial BH population by 12 Gyr, while the other models retained 1\% to 50\%.

The number of binary black holes (bottom panel of Figure~\ref{fig:nbh}) reaches a maximum after roughly $10\,$Myr due to formation of BH pairs via stellar evolution of primordial binaries. We assume an initial binary fraction of $5\%$ for all models, so this initial maximum scales simply with the total number of stars. The number of BBHs remains roughly constant from $t\approx 10\,$Myr until $t\approx100\,$Myr, when the black holes have mass segregated and formed a black hole subsystem. Once the black hole subsystem is formed, the number of BBHs decreases until reaching a value of roughly $1-10$ binaries, marking the onset of the ``black hole burning'' phase, in which the dynamical energy created by formation/hardening of black hole binaries roughly balances against the collapse of the cluster core \citep[for review, see][]{Kremer2020_bhburning}. As shown, this equilibrium value scales weakly with $N$, consistent with previous studies showing that the number of BBHs in a cluster is roughly independent of the total number of BHs at late times \citep[e.g.,][]{Chatterjee2017,Kremer2018a,MarinPinaGieles2024}.

\subsection{Black hole mergers}
\label{sec:hierarchical}

Throughout the 12 Gyr lifetime of the simulation, \texttt{colossus} produced $1{,}367$ binary BH mergers, 448 of which were hierarchical black hole mergers producing a BH of generation 3 or higher. In hierarchical mergers, each BH is considered generation 1 (1G) upon formation, and we define the generation of the merger product as the maximum generation of the two merger progenitors plus one. In addition, the merger itself is referred to by the maximum generation of the progenitors. For instance, a generation 2 black hole merging with a generation 1 black hole is known as a generation 2 merger, resulting in a generation 3 merger product.

\begin{figure*}
    \centering
    \includegraphics[width=\textwidth]{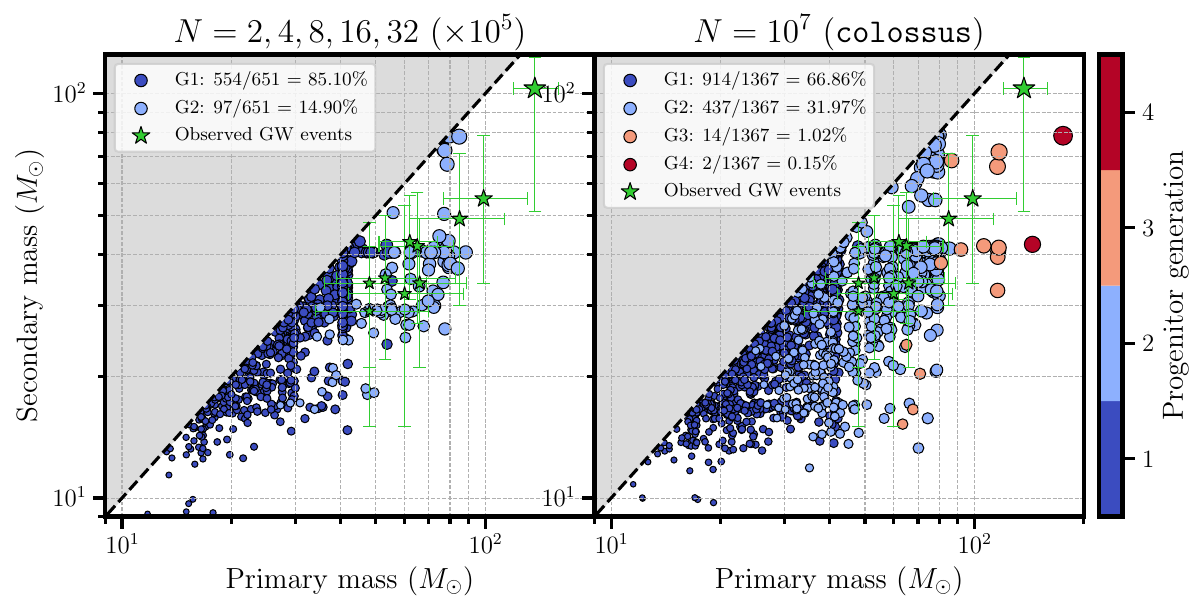}
    \caption{Primary vs. secondary mass for all binary black hole mergers in \texttt{colossus} (right panel), and an aggregate of five lower-mass models from the \texttt{CMC Catalog} with matching parameters except for initial mass (left panel). Each merger is colored by the higher generation of the two progenitor black holes, as described in the text. The ten current highest-mass GW observations from LVK are plotted in green, the three most massive of which lie exclusively within the high-generation G3+ merger space of \texttt{colossus}, including GW231123 (the upper-rightmost star in both panels). Not only do high-mass globular clusters like \texttt{colossus} produce many more mergers than lower-mass globular clusters, but high-generation mergers proportionally compose a much larger fraction of total mergers.
    \label{fig:m1m2}}
\end{figure*}

The primary vs. secondary mass for all binary BH mergers in \texttt{colossus} is shown in the right panel of Figure~\ref{fig:m1m2}, colored according to merger generation. For comparison, plotted in the left panel are all BBH mergers from the same five lower-mass models examined in Figure~\ref{fig:nbh}. Generation 2 mergers account for about 15\% of all BBH mergers in the five-model aggregate, but compose over 30\% of mergers in \texttt{colossus}. Furthermore, the five combined models did not produce any generation 3 mergers, whereas \texttt{colossus} alone produced fourteen generation 3 mergers and two generation 4 mergers. This is primarily due to the deeper potential well of \texttt{colossus}, which allows more merger products to be retained despite GW recoil kicks in excess of $100\,\text{km/s}$. These retained merger products can then potentially merge again after sinking back into the central black hole subcluster. The central escape velocity of a cluster can be estimated as 
\begin{equation}
    v_{\rm esc} = \sqrt{\frac{4GM_{\rm cl}}{r_h}} \approx 66\left(\frac{M_{\rm cl}}{10^6 M_\odot}\right)^{1/2} \left(\frac{r_h}{4\,\text{pc}}\right)^{-1/2}\,\text{km/s},
    \label{eq:binney}
\end{equation}
where $M_{\rm cl}$ is the total cluster mass and $r_h$ is the half-mass radius \citep[e.g.,][]{BinneyTremaine2008}. We adopt $r_h=4\,\text{pc}$ as a typical value for most of the cluster's lifetime (see Figure~\ref{fig:radius}), and show corresponding $v_{\rm esc}$ values on the secondary y-axis in the top panel of Figure~\ref{fig:nbh}. The escape velocity of \texttt{colossus} is near $150\,\text{km/s}$ for most of its lifetime, while all other models have $v_{\rm esc} \lesssim 100\,\text{km/s}$. This explains the discrepancy in high-generation mergers: in the lower-mass models, merger products are easily ejected by GW recoil kicks (typically $\approx 100\,\text{km/s}$), preventing hierarchical growth beyond one or two successive mergers. On the other hand, the deeper potential well of \texttt{colossus} and higher escape velocity threshold enable extended chains of hierarchical growth.

We note that GW recoil kick values, and therefore high-generation merger rates, are sensitive to the initial formation spins of 1G black holes. As $v_{\rm kick}$ values generally increase with the spins of merger progenitors, our assumption of $\chi_{\rm birth} = 0$ for 1G black holes produces optimistic high-generation merger rates \citep[for further discussion, see e.g.,][]{Rodriguez2019}. To approximate the effect of nonzero $\chi_{\rm birth}$ on the 2G+ merger rate, we can recalculate $v_{\rm kick}$ values for all 1G mergers in each 2G merger tree assuming $\chi_{\rm birth} > 0$, and consider the 2G merger suppressed if $v_{\rm kick} > v_{\rm esc}$ for any constituent merger. Assuming $\chi_{\rm birth} = 0.1$ ($0.2$), we find that 44\% (76\%) of 2G mergers in \texttt{colossus} are suppressed, demonstrating the strong impact of moderate $\chi_{\rm birth}$ on the high-generation merger rate. However, the angular momentum transport prescriptions of \cite{FullerMa2019spin} predict $\chi_{\rm birth} \sim 10^{-2}$ for most stellar evolution outcomes, so these suppression rates should be interpreted as upper limits. Namely, we find less than 8\% 2G suppression for $\chi_{\rm birth} = 0.05$, indicating our $\chi_{\rm birth} = 0$ assumption remains accurate to first order.

\begin{figure*}
    \centering
    \includegraphics[width=\textwidth]{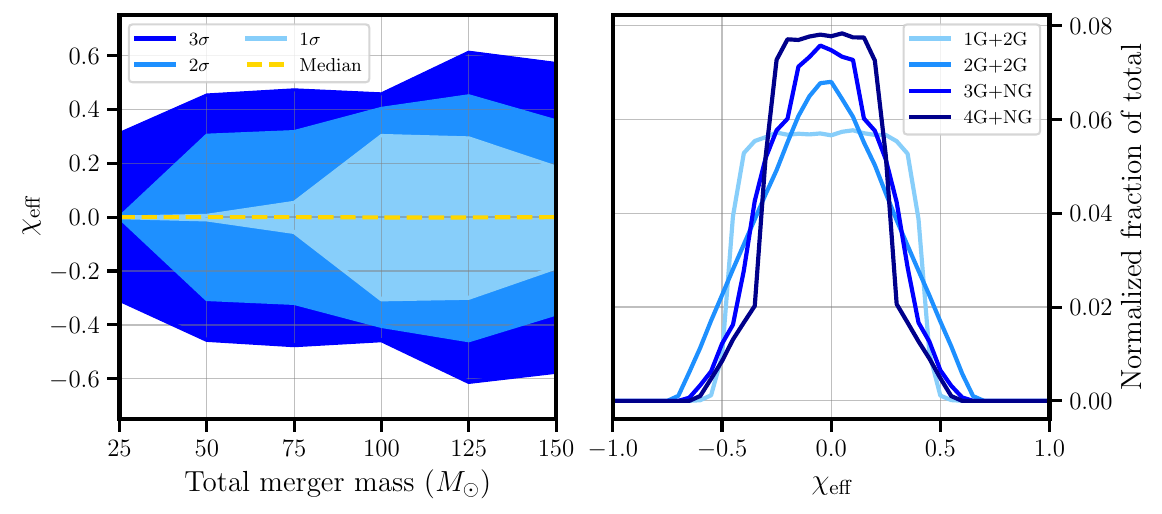}
    \caption{Net $\chi_{\rm eff}$ distributions for all BBH mergers in \texttt{colossus}, versus total merger mass (left panel) and hierarchical generation (right panel). $\chi_{\rm eff}$ distributions for individual BBH mergers were obtained through $10^3$ random angle draws on $\theta_i$ (Equation~\ref{eq:spin}), which were then pooled into corresponding bins based on total mass and hierarchical generation. Since all 1G+1G mergers have $\chi_{\rm eff} = 0$, the net $\chi_{\rm eff}$ distribution at $< 75~M_\odot$ is tightly centered about $\chi_{\rm eff} = 0$. Between $75~M_\odot$ and $125~M_\odot$, nearly all mergers involve at least one 2G black hole ($\chi \approx 0.7$), significantly broadening the $\chi_{\rm eff}$ distribution. Past $125~M_\odot$, many mergers involve a 3G or 4G black hole (which typically have lower spins than 2G), resulting in smaller $\chi_{\rm eff}$.}
    \label{fig:spin}
\end{figure*}

In addition, alongside each model's BBH mergers, Figure~\ref{fig:m1m2} also plots the ten most massive BBH GW sources observed to date (confident events only), as reported by the Gravitational Wave Open Science Center (GWOSC).\footnote{\url{https://gwosc.org}} The hierarchical merger history of these observed GW events, if any, is unknown, but their locations within Figure~\ref{fig:m1m2} suggest a merger generation of at least two, if they are indeed hierarchical. Pulsational pair-instability mechanisms within massive stars are expected to place an upper limit of roughly $40-50~M_\odot$ on first-generation black holes born via massive star collapse \citep[e.g.,][]{FowlerHoyle1964,Barkat1967,Fryer2001,Woosley2017}. In the \texttt{CMC} models described here, we assume stars that undergo pulsational-pair instability supernovae yield black holes of mass precisely $40.5\,M_{\odot}$, following the prescriptions of \citet{Belczynski2016}. In this case, ignoring contributions from major stellar collisions \citep{Kremer2020}, the highest mass a generation 2 black hole can have in a \texttt{CMC} model is approximately $80~M_\odot$. This limit can be observed in Figure~\ref{fig:m1m2} as a primary mass cutoff of about $80~M_\odot$ for generation 2 mergers. Since three of the most massive observed GW events plotted in Figure~\ref{fig:m1m2} have a primary mass exceeding $80~M_\odot$, these mergers would need to be generation 3 or higher \citep[assuming stellar collisions or accretion do not contribute significantly, see][for further discussion]{Kiroglu2025c}. Since generation 3 mergers are demonstrably rare in low-mass GCs, these observed GW events are much more likely to come from a GC with mass similar to \texttt{colossus}, rather than a typical lower-mass GC.

In addition to component masses, the observable parameter $\chi_{\rm eff}$ (effective spin) provides another avenue of comparison between BBH mergers in \texttt{colossus} and observed GW events. $\chi_{\rm eff}$ encodes the mass-weighted alignment between orbital angular momentum and component spin vectors, and is given by
\begin{equation}
    \chi_{\rm eff} = \frac{M_1 \chi_1 \cos\theta_1 + M_2 \chi_2 \cos\theta_2}{M_1 + M_2} \in (-1,1)
    \label{eq:spin}
\end{equation}
where $M_i$ and $\chi_i$ denote the mass and spin of a component BH, and $\theta_i$ denotes the angle between spin vector and orbital angular momentum. For BBH systems formed in dynamical environments, spin is generally uncorrelated with orbital angular momentum, so $\theta_i$ follows an isotropic distribution.\footnote{However, see \citet{Kiroglu2025b} for discussion of ways preferentially-aligned BBHs may form via stellar collisions in dynamical environments.} Therefore, for each BBH merger, we can obtain a $\chi_{\rm eff}$ distribution through random angle draws on $\theta_1$ and $\theta_2$. The combined $\chi_{\rm eff}$ distribution of all BBH mergers in \texttt{colossus} is shown in Figure~\ref{fig:spin}, versus total merger mass (left panel) and hierarchical generation (right panel). Since first-generation stellar black holes are assumed to form with birth spin of $\chi = 0$ \citep{FullerMa2019spin}, $\chi_{\rm eff} = 0$ for all 1G+1G mergers, which compose the majority of BBH mergers with total mass $\lesssim 75~M_\odot$. This can be observed in the left panel of Figure~\ref{fig:spin} as a tight distribution about $\chi_{\rm eff}=0$ for mergers with total mass $< 75~M_\odot$, and while not plotted in the right panel, it would appear as a delta function centered about $\chi_{\rm eff} = 0$. In contrast, any BBH merger involving a second-generation black hole (typically $\chi \approx 0.7$) will likely have a nonzero $\chi_{\rm eff}$ parameter, broadening the $\chi_{\rm eff}$ distribution between total masses of $75~M_\odot$ and $125~M_\odot$. This is qualitatively consistent with the trend observed in \cite{Antonini2025ppi}, which finds a tight $\chi_{\rm eff}$ distribution about $\chi_{\rm eff} = 0$ for mergers with primary mass $M_1 < \tilde{m} = 47.5^{+12.2}_{-8.6}M_\odot$, and a broader distribution past $M_1 > \tilde{m}$. Finally, given that the characteristic 2G $\chi = 0.7$ value is closer to the extremal value of $\chi=1$ than $\chi=0$, third-generation remnants are more likely to have a lower spin than their 2G progenitors after a randomly-oriented merger (assuming roughly comparable component masses). Therefore, past $125~M_\odot$ in which many mergers involve a 3G or 4G component, the $\chi_{\rm eff}$ distribution slightly compresses towards smaller values.

\begin{figure*}
    \centering
    \includegraphics[width=\textwidth]{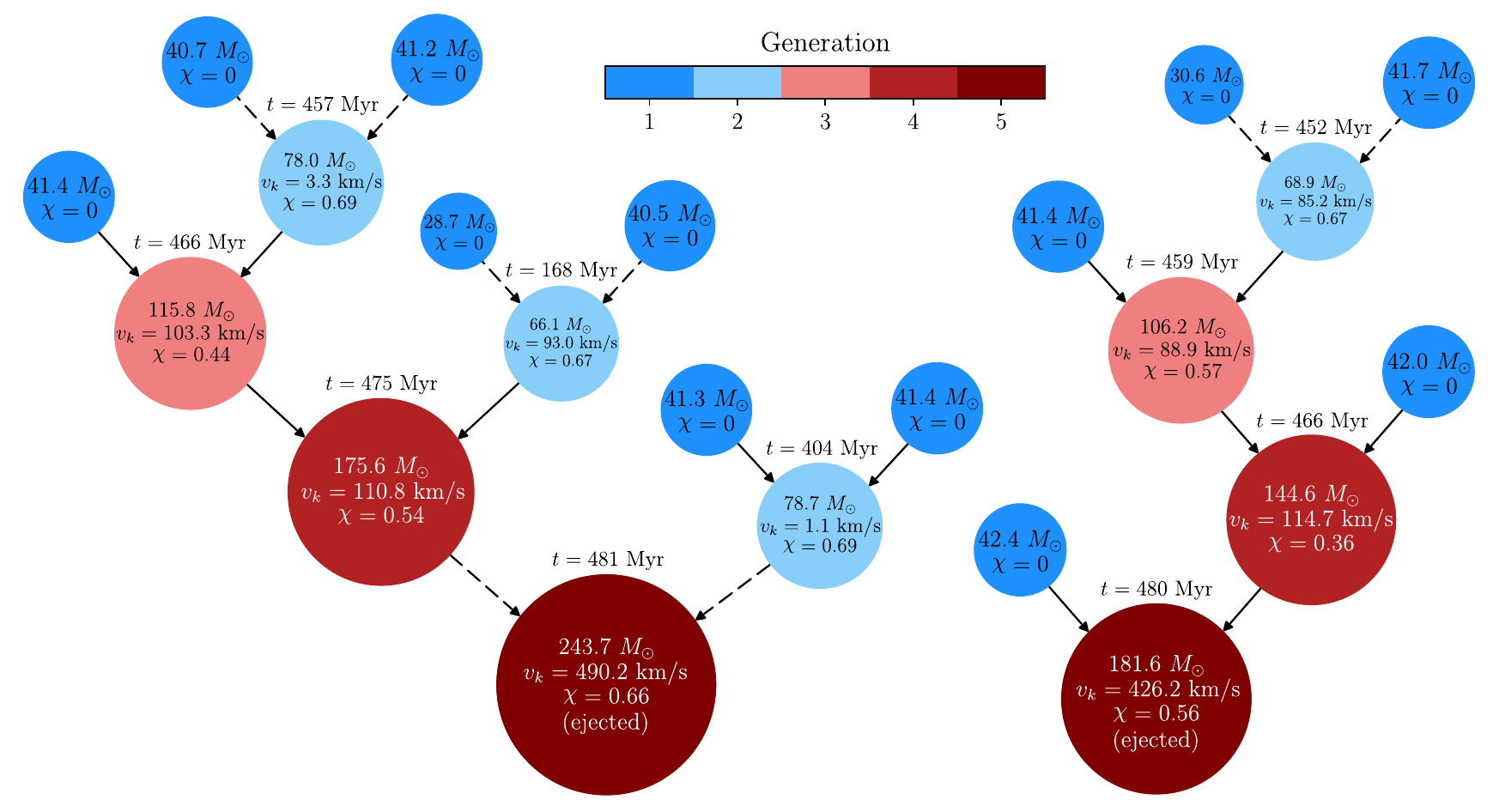}
    \caption{Hierarchical merger trees of the two highest-mass black hole in \texttt{colossus}, both generation 5 black holes with masses $243.7\:M_\odot$ and $181.6\:M_\odot$. Each black hole is colored according to its hierarchical generation, as described in the text. The mass, spin ($\chi$), GW kick velocity ($v_k$, applicable for generation 2+ BHs only), and merger time of each black hole is also listed. Solid lines indicate 2-body inspiral mergers, and dashed lines indicate 3-body capture mergers. Many BH mergers within both merger trees resulted in kick velocities of over 100~km/s, and would have been ejected from a typical lower-mass GC with $v_\text{esc} \lesssim 50~\text{km/s}$, but were retained in \texttt{colossus} ($v_\text{esc} \approx 150~\text{km/s}$).}
    \label{fig:familytree}
\end{figure*}

\begin{figure*}
    \centering
    \includegraphics[width=0.9\textwidth]{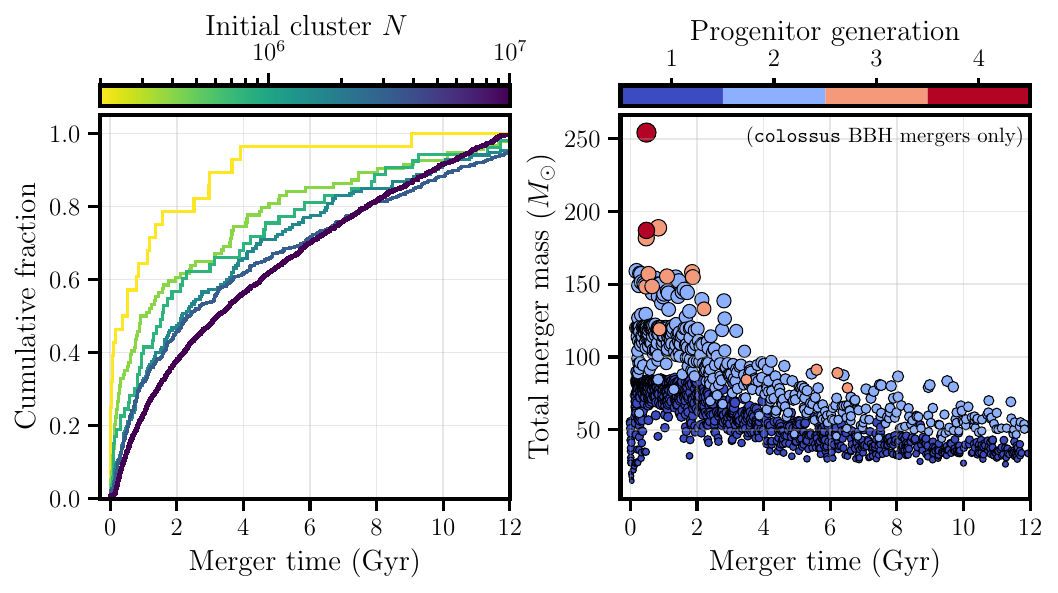}
    \caption{Left panel: Binary black hole merger delay time distributions for varying cluster masses, adopting the same models and color scheme as Figure~\ref{fig:nbh}. In general, more massive clusters feature longer delay times, a result of their longer relaxation times. Right panel: Total BBH mass versus merger time for all mergers occurring in \texttt{colossus}. Each merger is colored by its progenitor generation, as in Figure~\ref{fig:m1m2}. More massive BBHs have shorter mass segregation timescales, larger cross sections for dynamical interactions, and shorter GW inspiral times. Thus, more massive BBHs generally merge at earlier times compared to lower-mass BBHs.}
    \label{fig:delaytime}
\end{figure*}

\texttt{colossus} produced two generation 4 mergers: one 4G+2G and one 4G+1G merger, resulting in 5G remnant black holes of mass $243.7~M_\odot$ and $181.6~M_\odot$, respectively. After merging, both black holes received GW kicks in excess of 400~km/s and were ejected from the cluster, preventing any further hierarchical growth. The hierarchical merger histories of both black holes are shown in Figure~\ref{fig:familytree}, with GW kick velocities listed for generation 2 black holes and higher. Several mergers within both black holes' merger histories resulted in GW kicks exceeding 100~km/s, demonstrating why extended hierarchical growth in a typical GC ($v_\text{esc} \approx 50~\text{km/s}$) is extremely rare. \texttt{colossus}, however, with $v_\text{esc} \approx 150~\text{km/s}$ (see Figure~\ref{fig:nbh}), was able to retain these merger products, enabling further hierarchical growth.

The $243.7~M_\odot$ fifth-generation black hole is particularly noted for its similarity in mass to the recent event GW231123, which has total mass $225^{+26}_{-43}~M_\odot$ \citep{GW231123}. Recent work in \cite{PassengerGW23hierarchical} placed low probability on a 2G+2G merger being responsible for GW231123, and we find this conclusion consistent with both \texttt{colossus} and lower-mass models. The highest-mass 2G+2G remnant in \texttt{colossus}, being $153.5~M_\odot$, lies well below the total mass of GW231123. On the other hand, the 3G+ mergers exclusive to GCs as massive as \texttt{colossus} are more comparable to the mass of GW231123. This suggests that the most massive GCs (comparable to masses of nuclear star clusters) indeed provide a channel for high-mass GW observations, via 3G+ mergers \citep[e.g.][]{AntoniniRasio2016, FragioneRasio2023, GW231123}. However, from the perspective of spins, we find the highest-mass mergers formed via 3G+ mergers in \texttt{colossus} generally have spin values lower than the $\chi_1=0.90^{+0.10}_{-0.19}$ and $\chi_2=0.80^{+0.20}_{-0.51}$ values quoted for GW231123 at $90\%$ confidence (for reasons described in Figure~\ref{fig:spin}). In this case, spins may pose a challenge for formation of GW231123-like events via 3G+ mergers \citep[for further discussion, see][]{PassengerGW23hierarchical}. Alternative formation channels that may produce high spins include accretion in active galactic nuclei disks \citep[e.g.,][]{Tagawa2020,Vajpeyi2022,Bartos2025} or from stars in dense clusters \citep[e.g.,][]{Kiroglu2025c},
chemically homogeneous evolution \citep[e.g,][]{Marchant2016,Marchant2024,Stegmann2025}, and tidal spin up in stellar binaries \citep[e.g.,][]{Bavera2021,Ma2023,Qin2023}.

The two 4G mergers shown in Figure~\ref{fig:familytree} occur at times $481$ and $480\,$Myr, respectively. Assuming \texttt{colossus} was born $12\,$Gyr ($8\,$Gyr) ago, these merger times would correspond to redshift of roughly $z=2.96$ ($z=0.93$). We illustrate this point in the right panel of Figure~\ref{fig:delaytime}, which plots total mass versus merger time for all BBH mergers in \texttt{colossus}. Because of their shorter mass segregation timescales, larger cross sections for dynamical encounters (and thus faster hardening), and shorter GW inspiral times, higher-mass BBH systems merge earlier in the cluster's lifetime. In particular, all BBH mergers in \texttt{colossus} with total mass $>150~M_\odot$ occur within the first 2 Gyr of the cluster's lifetime. The median merger time of all 3G+ mergers in \texttt{colossus} is $985\,$Myr, corresponding to median redshift $z=2.49$ ($z=0.82$) for a cluster age of $12\,$Gyr ($8\,$Gyr). In this case, the oldest clusters in the local Universe may face challenges in forming massive hierarchical events similar to those currently seen by LVK, but these events could naturally be produced in more moderately-aged systems.

In the left panel of Figure~\ref{fig:delaytime} we show the delay time distribution for all mergers in \texttt{colossus} compared to low-mass \texttt{CMC} models (see Figure~\ref{fig:nbh} and Section~\ref{sec:properties}). As shown, more massive clusters generally feature longer delay times, simply because more massive clusters have longer relaxation times \citep[$t_{\rm rlx} \propto M_{\rm cl}^{1/2}$; e.g.,][]{BinneyTremaine2008}. However at late times ($t \gtrsim 8\,$Gyr), we see the delay time distribution for \texttt{colossus} steepens relative to lower-mass models. This occurs because unlike lower-mass models, nearly all of the mergers in \texttt{colossus} occur inside the cluster, which in general have shorter GW inspiral times relative to ejected mergers \citep[e.g.,][]{Rodriguez2018,Kremer2026_review}. We address this point further in Section~\ref{sec:properties}. 

In Table~\ref{table:merger_data}, we list the number of BBH mergers of various types in the \texttt{colossus} model. ``Ejected'' mergers refer to BBHs that merge following dynamical ejection from the cluster, ``2-body insp.'' mergers refer to BBHs that merge inside their host cluster following dynamical hardening and GW inspiral, and ``2-body cap.'', ``3-body cap.'', and ``4-body cap.'' refer to in-cluster mergers that occur via GW capture during single-single, binary-single, and binary-binary encounters respectively. For further discussion of the distinction between these various channels, see \citet{kremer2020modeling}.

\begin{deluxetable*}{cccccccccccc}
\tabletypesize{\footnotesize}
\tablewidth{0pt}
\tablecaption{Mergers and collisions in the $N=10^7$ \texttt{colossus} simulation\label{table:merger_data}}
\tablehead{
    \multicolumn{12}{c}{} \\
    \multicolumn{12}{c}{Compact Object Mergers}
}
\startdata     
\multicolumn{2}{c|}{Num. black holes} & \multicolumn{6}{c}{Black hole mergers} & \multicolumn{4}{|c}{Hierarchical mergers}\\
$t=1\,$Gyr & \multicolumn{1}{c|}{$t=12\,$Gyr} & Total & Ejected & 2-body insp. & 2-body cap. & 3-body cap. & 4-body cap. & \multicolumn{1}{|c}{1G+2G} & 2G+2G & 3G+NG & 4G+NG \\
13,499 & \multicolumn{1}{c|}{9,032} & 1,367 & 321 & 712 & 152 & 148 & 34 & \multicolumn{1}{|c}{374} & 63 & 14 & 2 \\
\hline
\hline
\multicolumn{2}{c|}{Num. neutron stars} & \multicolumn{3}{c|}{Neutron star mergers} & \multicolumn{7}{c}{}\\
$t=1\,$Gyr & \multicolumn{1}{c|}{$t=12\,$Gyr} & Total & Ejected & \multicolumn{1}{c|}{2-body insp.} & \multicolumn{7}{c}{n/a}\\
6,344 & \multicolumn{1}{c|}{6,250} & 0 & 0 & \multicolumn{1}{c|}{0} & \multicolumn{7}{c}{} \\
\hline
\hline
\multicolumn{2}{c|}{Num. white dwarfs} & \multicolumn{3}{c|}{White dwarf mergers} & \multicolumn{7}{c}{}\\
$t=1\,$Gyr & \multicolumn{1}{c|}{$t=12\,$Gyr} & Total & Ejected & \multicolumn{1}{c|}{2-body insp.} & \multicolumn{7}{c}{n/a}\\
304,829 & \multicolumn{1}{c|}{1,064,805} & 404 & 0 & \multicolumn{1}{c|}{404} & \multicolumn{7}{c}{} \\ 
\hline
\hline
\multicolumn{12}{c}{} \\
\multicolumn{12}{c}{Other Stellar Collisions} \\
\hline
~ & ~ & Star+Star & BH+star & NS+star & WD+star & WD+WD & NS+WD & BH+WD & ~ & ~ & ~\\
~ & ~ & 17,908 & 759 & 16 & 1,036 & 27 & 0 & 2 & ~ & ~ & ~ \\
\enddata
\end{deluxetable*}

\subsection{\texttt{Colossus} versus \texttt{Behemoth}}

The most comparable \texttt{CMC} model run to-date is the \texttt{behemoth} model of \citet{Rodriguez2020_SSC}, which was computed as part of the ``Great Balls of FIRE'' project designed to perform cluster simulations with initial conditions and time-dependent tidal forces extracted from a cosmological simulation \citep{Grudic2023,Rodriguez2023,Bruel2024A}. The \texttt{behemoth} model also contained roughly $10^7$ stars at birth, but featured two key differences relative to our \texttt{colossus} model: (1) \texttt{behemoth} was run for only $4\,$Gyr before it was fully disrupted by external tidal forces in its host galaxy and (2) \texttt{behemoth} is higher metallicity (roughly $0.5\,Z_{\odot}$), which translates to much lower mass black holes due to the influence of metallicity-dependent stellar mass loss due to winds \citep{Vink2001}. For example, the median primary mass for all first-generation black hole mergers in \texttt{behemoth} is $16.7\,M_\odot$, while it is $30.3\,M_\odot$ for \texttt{colossus}. From a numerical perspective, clusters with higher-mass black holes are computationally slower as the black holes undergo deeper collapse episodes driven by enhanced mass segregation \citep[e.g.,][]{kremer2020modeling,Rodriguez2022}. This in turn leads to shorter time steps compared to higher-metallicity simulations, which results in longer run times. Nonetheless, the \texttt{behemoth} model features similar conclusions to those drawn from our new \texttt{colossus} model, namely that the vast majority of BBH mergers occur inside the cluster and that extended chains of hierarchical mergers are enabled by the relatively higher cluster escape velocity \citep[for further detail and discussion of comparison to specific LVK events, see][]{Rodriguez2020_SSC}.

\subsection{Other transient events}

Although we focus here primarily on BBH mergers, for completeness, we also include in Table~\ref{table:merger_data} the number of other types of mergers and collisions occuring in the \texttt{colossus} simulation. Although roughly $6{,}000$ neutron stars are formed and retained in the cluster via stellar evolution, \textit{zero} binary neutron star mergers occur throughout the simulation. This is consistent with findings of previous studies \citep[notably][]{Ye2019} that have shown binary neutron star mergers are quite rare in clusters that have not yet reached cluster core collapse. We identify $404$ total white dwarf binary mergers in the simulation, all of which merge inside the cluster. This is also consistent with expectations from previous studies \citep[e.g.,][]{Kremer2021_wd} that have shown (1) white dwarf mergers preferentially occur inside their host cluster relative to BBH mergers, owing to their relatively small masses and correspondingly small semi-major axis values required for dynamical ejection and (2) similar to neutron star mergers, white dwarf mergers are rare relative to BH mergers in non-core-collapsed clusters.

We also list at the bottom of Table~\ref{table:merger_data} counts of other types of collisions. Notably, we find $759$ BH+star collisions occur over in the complete simulation. For comparison, $85$ and $37$ BH+star collisions occur in the corresponding \texttt{CMC Cluster Catalog} models with $N=32\times10^5,r_v=2\,$pc and $N=16\times10^5,r_v=2\,$pc, respectively. These BH+star collisions may give rise to bright electromagnetic transients powered by accretion of the disrupted star onto the BH \citep{Kremer2019_tde,Kremer2022_sph,Kremer2023_tde,Kiroglu2023}, potentially similar to transients events observed recently in the outskirts of nearby old massive elliptical galaxies \citep{Nicholl2023}. Similar to BBH mergers, we find there is a strong preference for BH+star collisions to occur in the most massive GCs, which may aid in potential follow-up of observed events as more massive GCs are brighter.

\begin{figure*}
    \centering
    \includegraphics[width=\textwidth]{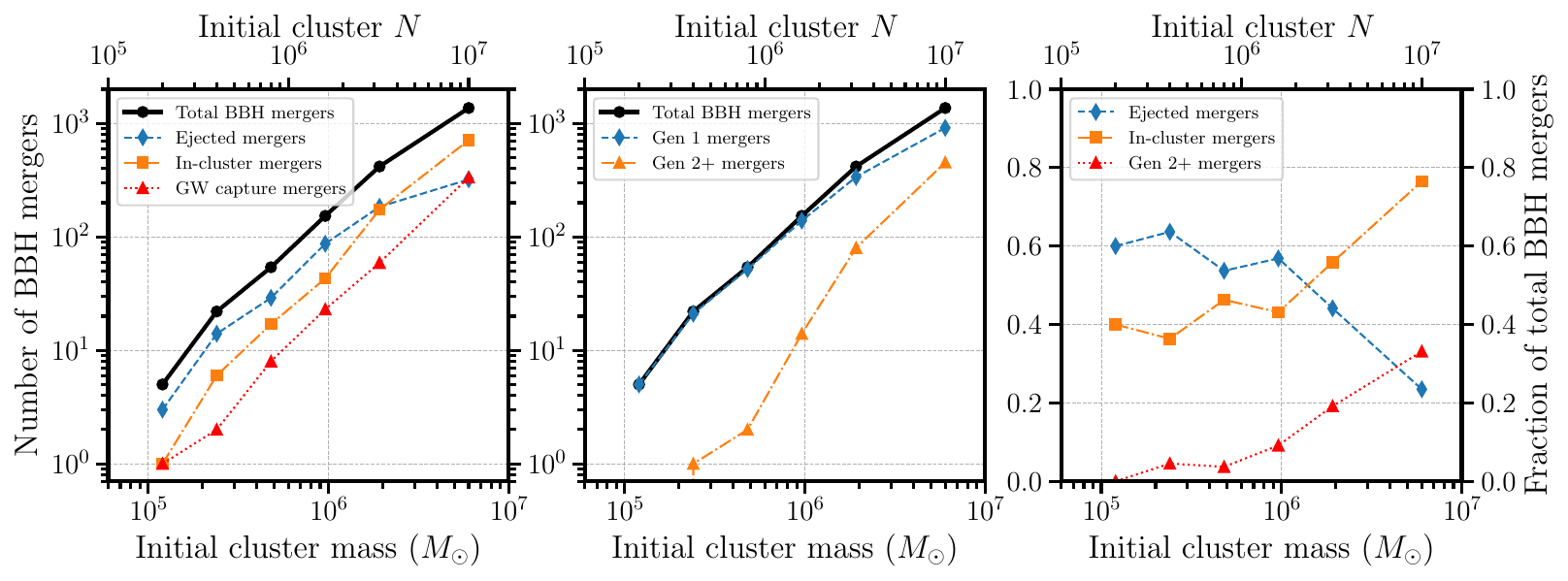}
    \caption{Distribution of binary black hole mergers versus initial cluster mass, sorted by merger channel (left panel) and hierarchical generation (center panel), also shown as a fraction of total mergers (right panel). Mergers of all channels increase with cluster $N$, except for ejected mergers, which begins to decrease in fraction after $N=32 \times 10^5$. Higher-mass clusters retain more in-cluster mergers, which can then merge again to form higher-generation mergers.}
    \label{fig:mergercount}
\end{figure*}

\section{The role of cluster properties}\label{sec:properties}

In this section, we explore how the number of BBH mergers varies with cluster properties, combining results from our new \texttt{colossus} model with the previous models of the \texttt{CMC Cluster Catalog}.

\subsection{Scaling of BBH mergers with cluster mass}\label{sec:bbh}

For comparison to \texttt{colossus}, we utilize the same five lower-mass \texttt{CMC Cluster Catalog} models plotted in previous figures, with the same initial conditions except for initial cluster $N$. In Figure~\ref{fig:mergercount}, we plot the number of BBH mergers in all six models versus initial cluster mass, $M_{\rm cl,0}$, and initial $N$, sorted by merger channel, hierarchical generation, and location. In the left panel of Figure~\ref{fig:mergercount}, we plot the total number of BBH mergers as a solid black curve, alongside three separate curves for the distinct merger channels \citep[for review, see][]{Kremer2026_review}: ejected mergers (blue diamonds), in-cluster 2-body inspirals (orange squares), and in-cluster GW capture mergers (red triangles). Note that GW captures here combines single-single, binary-single, and binary-binary capture channels listed in Table~\ref{table:merger_data}. A linear regression fit to the total BBH curve in log-log space yields the best-fit relation
\begin{equation}
    \label{eq:Nbbh}
    \log N_{\rm bbh} = (-6.38 \pm 0.47) + (1.42 \pm 0.08) \log (M_{\rm cl,0}/M_{\odot}).
\end{equation}

For comparison, \citet{AntoniniGieles2020} uses semi-analytic methods to find $N_{\rm bbh} \propto M_{\rm cl,0}^{1.6}$ for BBH mergers occurring at late times ($t > 8\,$Gyr). \citet{Hong2018} used Monte Carlo models performed with the \texttt{MOCCA} code \citep{Giersz2013} to find $N_{\rm bbh} \propto M_{\rm cl,0}^{1.3}$ for mergers occurring across all times. The $N_{\rm bbh} \propto M_{\rm cl,0}^{1.42}$ scaling found here is roughly consistent with these previous studies. We also performed a similar regression for present-day cluster mass $M_{\rm cl,f}$ (see Figure~\ref{fig:nbh}) and found a scaling of $N_{\rm bbh} \propto M_{\rm cl,f}^{1.36}$. Since the scaling with initial and final cluster masses are very similar, henceforth we simply adopt $N_{\rm bbh} \propto M_{\rm cl}^{1.4}$. Regression fits to other \texttt{CMC} models with varying $r_v$ show this scaling relation remains consistent, independent of $r_v$.

Both in-cluster and GW capture mergers roughly scale with the same proportions, but not ejected mergers, which begin to decrease in proportion between the $N=32 \times 10^5$ and $N=100 \times 10^5$ (\texttt{colossus}) models. This is simply due to the deeper potential well of higher-mass clusters, which makes it more difficult for objects to be ejected from the cluster. Theoretically, a sufficiently massive cluster would have all of its mergers occur in-cluster. This trend can also be observed in the right panel of Figure~\ref{fig:mergercount}, which instead plots ejected and in-cluster mergers as a fraction of total mergers. In \texttt{colossus}, roughly 80\% of all mergers occur inside the cluster, compared to about 40\% for the four lowest-mass models.

In the center panel of Figure~\ref{fig:mergercount}, we again plot the total number of BBH mergers versus initial cluster mass and $N$, as well as two curves for generation 1 (blue diamonds) and generation 2+ mergers (orange triangles). Below $N=16 \times 10^5$, roughly $90\%$ of all BBH mergers are generation 1, since the low mass of the cluster allows most BBH merger products to be immediately ejected via GW recoil kicks. For the $N=32 \times 10^5$ and $N=100 \times 10^5$ (\texttt{colossus}) models, generation 2+ mergers compose an increasing fraction of total mergers, up to 30\% for \texttt{colossus}. Generation 2+ mergers are also shown as a fraction of total mergers in the right panel of Figure~\ref{fig:mergercount} (red triangles). Linear regression fits to the \emph{total} generation 2+ curve in log-log space yields 
\begin{equation}
    \label{eq:highgen}
    \log N_{\rm 2G+} = (-10.95 \pm 0.97) + (2.02 \pm 0.16) \log (M_{\rm cl,0}/M_{\odot}).
\end{equation}
Of course, for sufficiently high cluster masses, this relation will break down since the number of generation 2+ mergers cannot exceed the total number of mergers in the cluster (Equation~\ref{eq:Nbbh}). However, within the parameter space of typical GC masses studied here ($M_{\rm cl} \lesssim 10^7\,M_{\odot}$), this $N_{\rm 2G+} \propto M_{\rm cl,0}^2$ scaling relation is expected to be appropriate. 

\subsection{Scaling of BBH mergers with half-light radius}

Similar to cluster mass, BBH merger populations are also expected to vary with cluster radius \citep[e.g.][]{kremer2020modeling}. We utilize 80 \texttt{CMC} models of varying initial $N$ and $r_v$ to derive a scaling relation in present-day cluster $r_h$, but exclude \texttt{colossus} in this analysis since no other model has comparable mass.

\begin{figure}
    \centering
    \includegraphics[width=\linewidth]{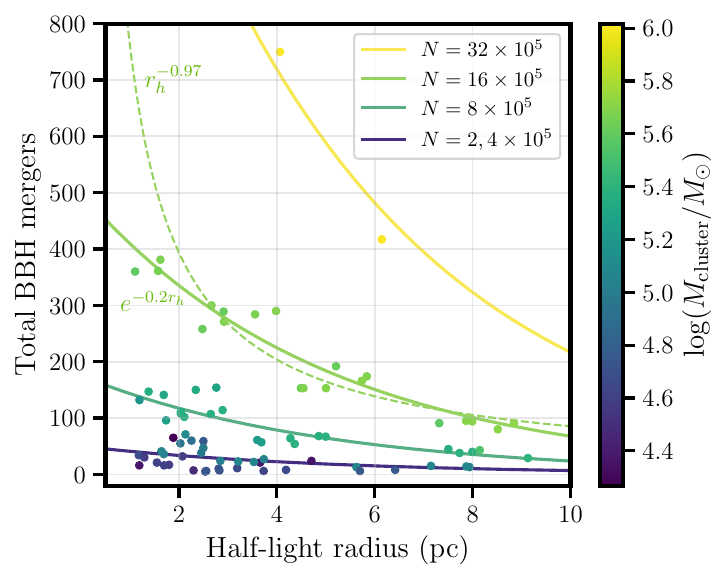}
    \caption{Total BBH mergers versus half-light radius for 80 \texttt{CMC} models with varying initial $N$ and $r_v$, along with the exponential fits given in Equation~\ref{eq:expfit}. The $N=2$ and $N=4$ models were grouped into one curve, since they did not exhibit sufficiently distinct BBH merger populations. The small number of $N=32 \times 10^5$ models made a regression fit impractical, so a simple scaling of the $N=16 \times 10^5$ curve is shown instead. All curves converged upon a decay constant of $\alpha = 0.2~\text{pc}^{-1}$, demonstrating $N_{\rm bbh} \propto e^{-0.2 r_h}$ regardless of cluster mass. A power-law fit to the $N=16 \times 10^5$ models of the form $N_{\rm bbh} \propto r_h^{-0.97}$ is shown as the green dotted curve, but this fit significantly overestimates $N_{\rm bbh}$ for the $r_v = 0.5\,\text{pc}$ models.}
    \label{fig:rhscaling}
\end{figure}

\begin{figure*}
    \centering
    \includegraphics[width=0.9\textwidth]{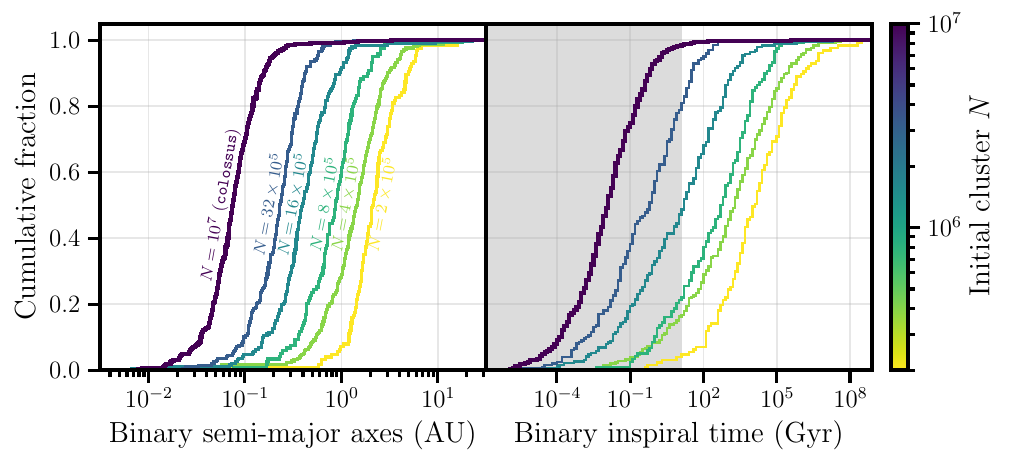}
    \caption{Semi-major axes and binary inspiral times for ejected BBH binaries in \texttt{colossus}, alongside the same five lower-mass models examined in Figure~\ref{fig:nbh}. The grey-shaded region in the right panel represents mergers with inspiral time of 12 Gyr or less. Higher-mass GCs eject tighter binaries with shorter inspiral times, especially \texttt{colossus}, in which over 98\% of ejected binaries have inspiral times under 12 Gyr.}
    \label{fig:esc}
\end{figure*}

As motivated by previous studies \citep{Hong2018,AntoniniGieles2020}, we initially attempted to fit a power-law correlation to $r_h$ of the form $N_{\rm bbh} \propto r_h^{-\alpha}$, but found that power-law fits were unable to produce accurate results for the $r_v=0.5~\text{pc}$ \texttt{CMC} models (previous studies did not consider $r_v < 1~\text{pc}$ clusters in their analyses). We were able to reproduce \citet{AntoniniGieles2020}'s $N_{\rm bbh} \propto r_h^{-0.9}$ relation by excluding the $r_v=0.5~\text{pc}$ models from the regression fit, yielding $N_{\rm bbh} \propto r_h^{-0.97}$ (plotted as a dashed curved in Figure~\ref{fig:rhscaling}), but this fit significantly overestimates $N_{\rm bbh}$ at low $r_h$. We instead utilize an exponential function of the form $N_{\rm bbh} \propto e^{-\alpha r_h}$ to better fit the data. The best-fit curves for the total BBH merger count with varying $r_h$ obtained through exponential regression are given in Equation~\ref{eq:expfit}, and plotted in Figure~\ref{fig:rhscaling}. The $N=2$ and $N=4$ models did not exhibit sufficiently distinct BBH merger populations, so they were grouped into one curve.
\begin{equation}
    \label{eq:expfit}
    N_{\rm bbh} =
    \begin{cases}
        489 \times e^{-(0.20 \pm 0.02) \times r_h}; & N=16 \times 10^5 \\
        184 \times e^{-(0.21 \pm 0.04) \times r_h}; & N=8 \times 10^5 \\
        45 \times e^{-(0.21 \pm 0.09) \times r_h}; & N=2,4 \times 10^5
    \end{cases}
\end{equation}
The $\alpha$ values of all three curves are very similar, so we henceforth adopt $\alpha = 0.2~\text{pc}^{-1}$ for all cluster masses.

We also performed exponential regression fits for generation 2+ mergers, yielding the following best-fit curves:
\begin{equation}
    \label{eq:exp2g}
    N_{\rm 2G+} =
    \begin{cases}
        136 \times e^{-(0.37 \pm 0.04) \times r_h}; & N=16 \times 10^5 \\
        184 \times e^{-(0.45 \pm 0.08) \times r_h}; & N=8 \times 10^5 \\
    \end{cases}
\end{equation}
The $N=2,4$ models did not have enough generation 2+ mergers to perform a regression analysis. We will simply assume $\alpha_{\rm 2G+} = 0.4~\text{pc}^{-1}$ for high-generation mergers, yielding $N_{\rm 2G+} \propto e^{-0.4 r_h}$.

\subsection{Ejected BBH properties}

In the dense environment of GCs, BBHs experience repeated close encounters with other single stars and binary systems. Each of these encounters transfer energy from the BBH's internal energy into the kinetic energy of the interacting objects, which leads to BBH hardening and dynamical recoil \citep[e.g.,][]{Rodriguez2016mergers,Kremer2026_review}. Energetics arguments show center-of-mass recoil velocity of a binary is comparable to its orbital velocity $v_{\rm orb} \approx \sqrt{Gm_{\rm bh}/a}$. Therefore, as a BBH hardens, it recoils more. The characteristic minimum separation a binary can achieve via dynamical hardening is set by its host cluster's escape velocity (Equation~\ref{eq:binney}). Once a BBH becomes sufficiently compact, its recoil velocity will exceed its host's escape velocity, and it will be ejected from the cluster. A subset of these ejected binaries will have sufficiently short GW inspiral times to ultimately merge within a Hubble time. Since higher-mass GCs have higher escape energy thresholds, ejected BBH systems of a higher-mass GC will have higher binding energies, shorter semi-major axes, and correspondingly shorter GW inspiral times \citep{Rodriguez2016mergers}.

We demonstrate this result in Figure~\ref{fig:esc}, showing cumulative histograms of binary semi-major axes and inspiral times for ejected BBH systems in \texttt{colossus}. Again, we plot the same quantities of five lower-mass models for comparison, as in previous figures. 98\% of ejected BBH systems in \texttt{colossus} have inspiral times less than a Hubble time, with the other models ranging from 4\% to 79\%.

This trend also explains why the majority (roughly 80\%) of the BBH mergers in \texttt{colossus} occur \textit{inside} the cluster (see Table~\ref{table:merger_data}). The GW inspiral time of a BBH scales steeply with semi-major axis \citep[$t_{\rm insp} \propto a^4$;][]{Peters1964}. This means BBHs in more massive clusters that are near the critical orbital separation for dynamical ejection are more susceptible to merging inside their host before ejection can occur \citep[see also, e.g.,][]{Rodriguez2019,Kremer2020,AntoniniGieles2020}. Indeed, this explains the trend shown in the right-panel of Figure~\ref{fig:mergercount} where the ratio of in-cluster to ejected mergers increases with cluster mass.

\section{Mapping to Virgo globular clusters}
\label{sec:virgo}

With the addition of \texttt{colossus}, the \texttt{CMC Cluster Catalog} now adequately spans the mass-$r_h$ parameter space of observed GCs in the Virgo Supercluster. This enables extrapolation of quantities like the total number of binary BH mergers and BBH merger rate from the \texttt{CMC} catalog to observed Virgo clusters, and prediction of total BBH merger rate from GC sources in the Virgo supercluster as a whole. We introduce a mass-$r_h$ distance function to map Virgo GCs to their closest \texttt{CMC} catalog counterparts, and utilize the scaling relations in cluster mass and $r_h$ derived in Section~\ref{sec:properties} to predict BBH merger quantities for each Virgo cluster, based on its closest \texttt{CMC} counterparts. Unless otherwise specified, all \texttt{CMC} model properties referred to in this section are present-day values ($t > 12\,\text{Gyr}$).

\subsection{Cluster mapping with distance function} \label{sec:mapping}

\begin{figure}
    \centering
    \includegraphics[width=\linewidth]{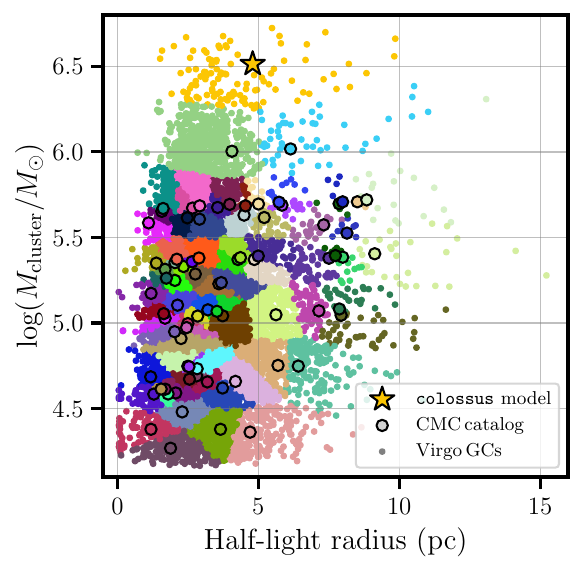}
    \caption{Mass-$r_h$ parameter space of observed Virgo GCs and \texttt{CMC} clusters. Each \texttt{CMC} cluster is given an arbitrary color, and every Virgo cluster is colored according to its closest \texttt{CMC} counterpart, with similarity quantified using the distance formula in Equation~\ref{eq:distance}. \texttt{colossus} encompasses the gold region at $\log (M/M_{\odot}) \gtrsim 6.3$, which previously did not have coverage in the \texttt{CMC Cluster Catalog}.}
    \label{fig:paramspace}
\end{figure}

For Virgo GC properties, we use the results from the ``ACS Virgo Cluster Survey'' \citep[e.g.,][]{Cote2004}. In particular, we use the catalog in \citet{Jordan2009} to attain present-day cluster masses and half-light radii.

\begin{figure*}
    \centering
    \includegraphics[width=\textwidth]{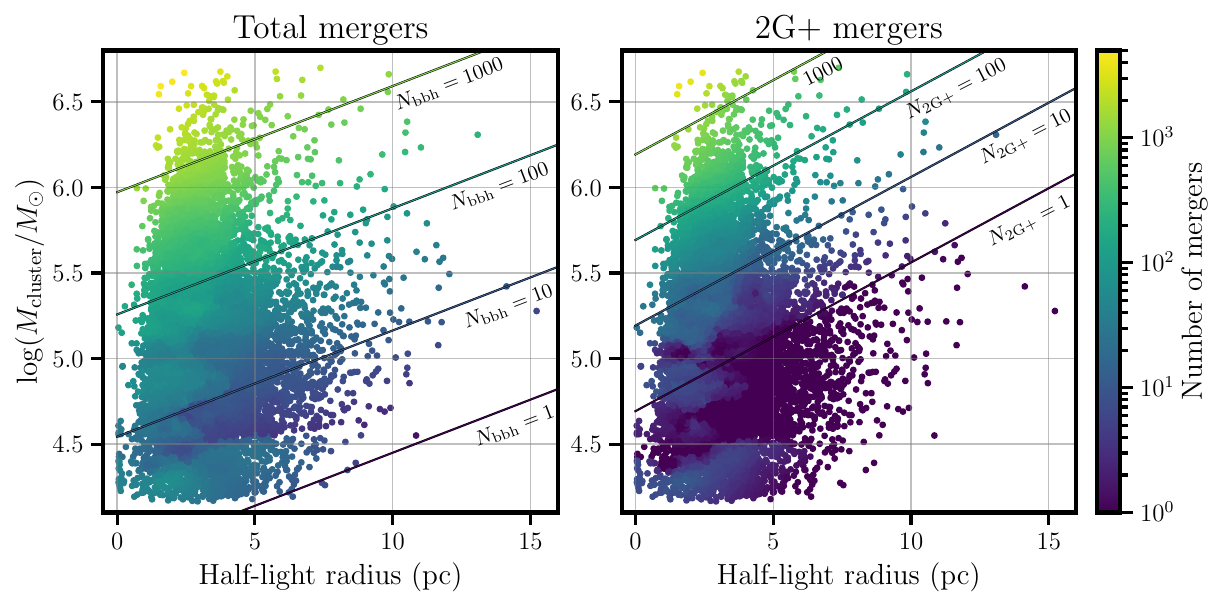}
    \caption{Estimated total BBH merger populations ($N_{\rm bbh}$, left panel) and generation 2+ merger populations ($N_{\rm 2G+}$, right panel) for each GC in the Virgo cluster, utilizing the scaling relations and weighting scheme described in the text. Constant curves for 1, 10, 100, and 1000 mergers are shown in both panels, obtained by scaling off of a middling \texttt{CMC} model with $\log (M_{\rm cluster}/M_\odot) = 5.7$, $r_h = 5.0\,\text{pc}$, $N_{\rm bbh} = 153$, and $N_{\rm 2G+} = 14$ \citep[model 104 in][]{kremer2020modeling}. As expected, merger counts generally grow with increasing cluster mass and decreasing $r_h$.}
    \label{fig:virgobbh}
\end{figure*}

Figure~\ref{fig:paramspace} compares the parameter space of Virgo GCs with the \texttt{CMC} catalog clusters in mass-$r_h$ space. For each Virgo cluster, its similarity to a \texttt{CMC} model in mass-$r_h$ space can be quantified using the distance function
\begin{equation}
    \label{eq:distance}
    \begin{aligned}
        d  & = (C \cdot \Delta \log M)^2 + (\Delta r_h)^2 \\
           & = \left(C\log \frac{M_{\texttt{CMC}}}{M_\text{Virgo}}\right)^2 + \left(r_h^{\texttt{CMC}} - r_h^{\text{Virgo}}\right)^2,
    \end{aligned}
\end{equation}
where $C=10$ represents a constant scaling factor on $\Delta \log M$ to account for the differing scales of $\log M$ and $r_h$. A similar constant scaling factor could be applied to $\Delta r_h$, but we did not find this necessary to achieve good coverage of the parameter space. In Figure~\ref{fig:paramspace}, each Virgo cluster is colored according to its closest \texttt{CMC} model, as quantified in Equation~\ref{eq:distance}.

\subsection{Interpolating the \texttt{CMC} model grid}

Using Equation~\ref{eq:distance} to quantify the similarity between \texttt{CMC} models and Virgo clusters, we can utilize the scaling relations of Section~\ref{sec:properties} to interpolate the results of the \texttt{CMC} models and predict the number of mergers for each observed Virgo GC, and also extrapolate to Virgo GCs outside of the region covered by the \texttt{CMC} model parameter space.

In Section~\ref{sec:properties}, we derived the scaling relations $N_{\rm bbh}~\propto~M_{\rm cl}^{1.4}$ (Equation~\ref{eq:Nbbh}) and $N_{\rm bbh}~\propto~e^{-0.2 r_h}$ (Equation~\ref{eq:expfit}) in cluster mass $M$ and half-light radius $r_h$. Since the mass relation is independent of half-light radius and vice-versa, we combine the two relations as a separable function of the form
\begin{equation}
    \label{eq:totalfit}
    N_{\rm bbh}(M_{\rm cl},r_h) \propto M_{\rm cl}^{1.4}e^{-0.2r_h}.
\end{equation} 
When using this combined scaling relation to predict $N_{\rm bbh}$ for an arbitrary Virgo GC, normalization is provided by the $N_{\rm bbh}$ count of a nearby \texttt{CMC} model. Therefore, $N_{\rm bbh}$ for any arbitrary Virgo GC can be extrapolated from a \texttt{CMC} model using Equation~\ref{eq:nbbhvirgo}:
\begin{multline}
    \label{eq:nbbhvirgo}
    N_{\rm bbh}^{\rm Virgo} = N_{\rm bbh}^{\texttt{CMC}} \times \left(\frac{M_{\rm Virgo}}{M_{\texttt{CMC}}}\right)^{1.4} \\ \times \exp\left[-0.2 \times (r_h^{\rm Virgo} - r_h^{\texttt{CMC}})\right]
\end{multline}
We can also estimate the number of generation 2+ hierarchical mergers in a Virgo cluster, $N_{\rm 2G+}$, utilizing the $N_{\rm 2G+} \propto M^2$ and $N_{\rm 2G+} \propto e^{-0.4r_h}$ relations from Section~\ref{sec:properties} to modify Equation~\ref{eq:nbbhvirgo}. This amounts to simply replacing the mass power with $2$ and the $r_h$ decay constant with $-0.4$, as well as $N_{\rm bbh}^{\texttt{CMC}}$ with $N_{\rm 2G+}^{\texttt{CMC}}$.

To prevent granular behavior of interpolated $N_{\rm bbh}^{\rm Virgo}$ counts, instead of simply scaling off of the single closest \texttt{CMC} model, we take a weighted average of the five closest models. Let $d_n$ denote the distance of the $n$-th closest \texttt{CMC} model in mass-$r_h$ space (Equation~\ref{eq:distance}), and let $N^{\rm bbh}_n$ denote the $n$-th closest model's extrapolated $N_{\rm bbh}^{\rm Virgo}$ value (Equation~\ref{eq:nbbhvirgo}). Then the final $N_{\rm bbh}^{\rm Virgo}$ count is given by
\begin{equation}
    \label{eq:nbbhweight}
    N_{\rm bbh}^{\rm Virgo} = \frac{\sum_{n=1}^5 (d_1 / d_n) \times N_n^{\rm bbh}}{\sum_{n=1}^5 (d_1/d_n)}.
\end{equation}
Using the weighting scheme of Equation~\ref{eq:nbbhweight}, a \texttt{CMC} model that is twice as far from the target Virgo GC as the closest model will be weighted half as much, thrice as far will be weighted one-third as much, etc. An equivalent method can be used to calculate $N_{\rm 2G+}^{\rm Virgo}$, simply by replacing $N_n^{\rm bbh}$ with $N_n^{\rm 2G+}$ (obtained using the appropriate 2G+ scaling relations). Since Equation~\ref{eq:nbbhvirgo} may break down when extrapolating over vast mass or $r_h$ ranges, only the five closest \texttt{CMC} models are considered, and Virgo GCs which exceed the initial mass of \texttt{colossus} are excluded from the rest of the analysis ($\log M_{\rm cl} > 6.8$, only six GCs in the \citet{Jordan2009} catalog).

\subsection{Predicted Virgo cluster BBH merger populations}

We utilize Equation~\ref{eq:nbbhweight} to calculate $N_{\rm bbh}$ and $N_{\rm 2G+}$ for each GC in the Virgo cluster. Figure~\ref{fig:virgobbh} again plots the mass-$r_h$ parameter space of Virgo GCs (excluding six GCs with mass exceeding the initial mass of \texttt{colossus}), colored according to the predicted number of total BBH mergers ($N_{\rm bbh}$, left panel) and the predicted number of generation 2+ hierarchical mergers ($N_{\rm 2G+}$, right panel). The total number of mergers in each panel is shown to increase as the cluster density increases, as expected. For very low-mass clusters with present-day masses less than roughly $10^{4.5}\,M_{\odot}$, the predicted number of mergers can be seen to increase slightly. This is because some of these clusters are best matched by \texttt{CMC} models that were initially more massive (and therefore formed many BBH mergers, especially at early times), but lost a significant amount of mass via enhanced tidal stripping in their galactic potential \citep[see][for further discussion]{kremer2020modeling}.

\begin{figure}
    \centering
    \includegraphics[width=\linewidth]{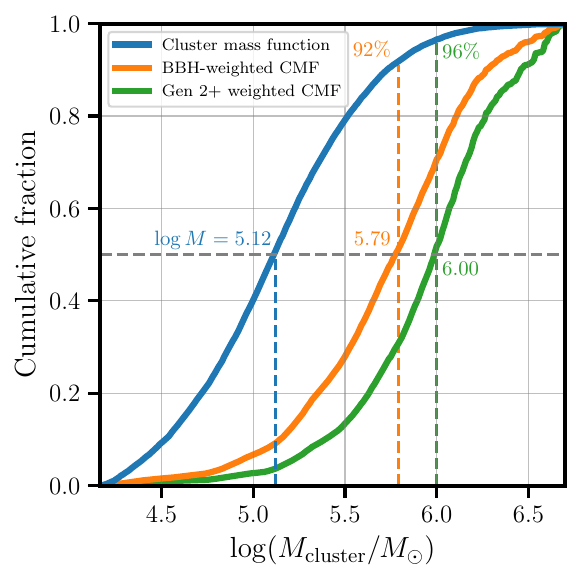}
    \caption{Cumulative cluster mass function (CMF) of Virgo GCs (blue), weighted by both total BBH merger populations (orange) and 2G+ merger populations (green). The 50\% thresholds for all three curves are also labeled. The heaviest 8\% of Virgo GCs are responsible for 50\% of all BBH mergers, and the heaviest 4\% are responsible for 50\% of all 2G+ mergers.}
    \label{fig:virgohist}
\end{figure}

In Figure~\ref{fig:virgohist}, we show the contribution of different parts of the cumulative cluster mass function to the overall BBH merger rate. As a blue curve, we plot the unweighted cumulative cluster mass function for all observed GCs in Virgo. As shown, the median GC mass is roughly $\log (M/M_{\odot}) = 5.12$. In orange and green, we plot the cumulative cluster mass function weighted by the predicted number of BBH mergers and the predicted number of 2G+ mergers, respectively, with the corresponding 50\% thresholds also shown. This indicates that $50\%$ of all BBH mergers formed in old GCs similar to the Virgo GC sample occur in clusters of mass $\log (M/M_{\odot}) = 5.79$ or higher, while $50\%$ of all 2G+ BBH mergers formed in GCs occur in clusters of mass in excess of $\log (M/M_{\odot}) = 6$. An alternative interpretation is that the upper $8\%$ ($4\%$) of the cumulative cluster mass function contributes to half of the full (2G+) BBH merger sample coming from GCs.

\section{Summary and conclusions}
\label{sec:summary}

\subsection{Summary}

In this study, we have presented the results from a 10-million-body globular cluster simulation, \texttt{colossus}, performed with the \texttt{CMC} code. The large initial mass of this cluster ($6 \times 10^6~M_\odot$) enables exploration of many trends across a wide range in cluster mass, such as black hole populations, binary black hole merger properties and quantities, delay time distributions, and parameters of ejected binary black hole systems. We identify the following key results:

\begin{enumerate}
    \item Combining our \texttt{colossus} model with lower-mass cluster models computed as part of the \texttt{CMC Cluster Catalog}, we find that the number of binary black hole mergers per cluster scales roughly as $N_{\rm bbh} \propto M_{\rm cl}^{1.4} e^{-0.2r_h}$. This is roughly consistent with results from previous studies derived from semi-analytic methods.
    
    \item This model empirically demonstrates the prevalence of high-generation hierarchical mergers in high-mass GCs, many of which have similar parameters to the most massive gravitational wave events detected by the LVK collaboration to date. We identify hierarchical merger chains extending up to fifth generation and black hole masses up to roughly $250\,M_{\odot}$, providing insight into the formation of intermediate-mass black holes and high-mass GW events through dynamical channels.
    
    \item We have developed a weighting scheme to map between our cluster simulations and the many thousands of GCs observed in the galaxies of the Virgo Supercluster. This tool enables prediction of the black hole merger history for realistic observed GCs across all masses and radii.
    
    \item As an example of this tool, we demonstrate that $50\%$ of binary black hole mergers occurring in old GCs form in clusters of mass $\log (M_{\rm cl}/M_{\odot}) = 5.79$ or more. This is significantly higher than the median GC mass of $\log (M_{\rm cl}/M_{\odot}) = 5.12$, confirming that the most massive clusters contribute an outsized influence upon the overall black hole merger rate.
\end{enumerate}

The output from \texttt{colossus} is now publicly available as part of the \texttt{CMC Cluster Catalog}.\footnote{\url{ https://cmc.ciera.northwestern.edu}} Additionally, the Virgo interpolation scheme developed in this study is available on GitHub,\footnote{\url{https://github.com/aidanmai/colossus}} along with all BBH merger data from \texttt{colossus}, and notebooks for generating the figures in Section~\ref{sec:virgo}.

\subsection{Future Work}

Continued exploration of this high-mass parameter space will be essential for fully understanding the role of dynamical processes in gravitational wave astronomy. For example, although our \texttt{colossus} model produces over 1,300 black hole mergers, only a handful of these reach fourth generation or higher and total masses in excess of $150\,M_{\odot}$. Additional models are necessary to resolve these small-number statistics and compare in detail to specific observed LVK events. Furthermore, massive clusters with initial virial radii both larger and smaller than \texttt{colossus} ($r_v=2\,$pc) are needed to more fully span the parameter space exhibited in observed clusters (see upper region of Figure~\ref{fig:paramspace}).

We have focused here on black hole growth via hierarchical mergers, however other growth channels are likely possible in dense stellar clusters. One scenario is growth via massive stellar mergers and collisions while the cluster is very young ($t\lesssim10\,$Myr). A number of studies have demonstrated these may provide a pathway for forming black holes in or beyond the pair-instability mass gap \citep[e.g.,][]{DiCarlo2019,Kremer2020,Gonzalez2021,Costa2022,Ballone2023}. Indeed, $767$ black holes with masses in the pair-instability gap formed via binary stellar collisions/mergers in \texttt{colossus} (see the small handful of G1 points in Figure~\ref{fig:m1m2} with primary mass $>40.5\,M_\odot$ which formed via this channel). Additionally, several studies have shown that black holes may also grow via accretion upon collision with other massive stars \citep[e.g.,][]{Giersz2015,Kiroglu2025a,Kiroglu2025c}. \citet{Rose2023} pointed out that this channel may be especially prominent in very dense star clusters, with densities similar to the Galactic center. In \texttt{colossus}, we identify $759$ black hole+star collisions events over the cluster's lifetime. For the vast majority of these ($90\%$), the $M_{\star}/M_{\rm bh}$ mass ratio is $0.1$ or less. In this regime, significant black hole growth is unlikely \citep{Kremer2022_sph}. Only five collisions feature $M_{\star}/M_{\rm bh}$ values of 0.5 or higher (all occurring before $25\,$Myr) where significant growth may be feasible. However, we have assumed in \texttt{colossus} a $5\%$ binary fraction for all stars. Realistic clusters may very well have much higher binary fractions, especially for massive stars \citep[e.g.,][]{Offner2023}. Higher binary fractions may significantly increase the number of massive star--black hole collisions \citep{Kiroglu2025a}. We reserve exploration of this possibility for future models.

A key recent result is the confirmation of an intermediate-mass black hole (IMBH) of mass roughly $10^4\,M_{\odot}$ in the Milky Way cluster $\omega$Cen \citep{Haberle2024}. It remains to be seen whether massive central black holes are a common feature of all massive GCs, or if $\omega$Cen is exceptional in this respect. In many clusters, it can be challenging to distinguish observationally between a single massive black hole and a central subsystem contain many stellar-mass black holes \citep[e.g.,][]{Vitral2023}. That being said, preliminary \texttt{CMC} models containing central IMBHs demonstrate that the long-term dynamics and black hole merger histories of such systems can be quite distinct from those of clusters like \texttt{colossus} \citep{GonzalezPrieto2025}. Future \texttt{CMC} studies will explore in detail how the presence (or lack thereof) of central IMBHs in massive GCs impacts our predictions for LVK merger events. 

\acknowledgements

The authors thank Carl Rodriguez and Fred Rasio for helpful discussions throughout the preparation of the manuscript. We also thank the anonymous referee for helpful feedback. A.M. acknowledges support from the Undergraduate Research Scholarship (URS) summer program at UCSD. F.K. acknowledges support from the CIERA Postdoctoral Fellowship.

\bibliographystyle{aasjournal}
\bibliography{mybib.bib}

\end{document}